\documentclass[aps,prb,twocolumn,showpacs]{revtex4}

\usepackage{amsmath}
\usepackage{bm} 
\usepackage{graphicx} 

\usepackage{gensymb}

\begin{document}

\title{Self-organizing structures in immiscible crystals}

\author{Marcin  Mi{\'n}kowski$^1$, Magdalena A. Za{\l}uska--Kotur$^1$,{\L}ukasz A.Turski$^2$,  and  Grzegorz Karczewski $^1$}
\email{minkowski@ifpan.edu.pl, zalum@ifpan.edu.pl,laturski@cft.edu.pl,karcz@ifpan.edu.pl} 
\address{$^1$Institute of Physics, Polish
Academy of Sciences, Al.~Lotnik{\'o}w 32/46, 02--668 Warsaw, Poland,\\
$^2$Center for Theoretical  Physics, Polish
Academy of Sciences, Al.~Lotnik{\'o}w 32/46, 02--668 Warsaw, Poland,}

\begin{abstract}
 Spinodal decomposition process in the system of  immiscible  PbTe/CdTe compounds is analyzed as an  example of a self-organizing structure. The immiscibility of the constituents leads to the observed morphological transformations like anisotropy  driven  formation of quantum dots and nanowires, and to the phase separation at the highest temperatures. Proposed  model  accomplishes bulk and surface  diffusion together with the  anisotropic  mobility of material components. We analyze   its properties by   kinetic Monte Carlo simulations and show that   it is able to reproduce  all  of the structures  observed experimentally in the process of   PbTe/CdTe growth.    We show that studied mechanisms of dynamic processes play different role in the creation of zero-- ,  one-- , two-- and finally three-dimensional structures. The shape of grown  structures    is   different for relatively thick multilayers when bulk diffusion cooperates with the anisotropic mobility, in annealed structure when isotropic bulk diffusion only decides about the process and finally  for thin multilayers when surface diffusion is the most decisive factor. We compare our results  with experimentally grown systems  and show that proposed  model explains the diversity of observed structures. 

\end{abstract}

\maketitle

\section{Introduction}
Low-temperature, layer--by--layer epitaxial growth techniques are commonly employed  for  creation of  complex layered nanostructures with atomically flat interfaces and vanishing material intermixing at the interfaces. However, in some cases  growth temperature or the after--growth annealing may result in a morphological reorganization of nominally layered structures adopting their  shape to the given thermodynamic conditions. For instance, such  case is represented by nanoscale superlattices made of immiscible materials. Thermally driven morphological transitions observed in immiscible superlattices can be regarded as an analog of the spinodal decomposition process which proceeds in the solid phase on a nanometer scale. Self-organization of layered structures in the spinodal decomposition process has been observed in many  immiscible material systems \cite{deGennes,cahn,binder,liu,vollmer,dietl}.

Spinodal decomposition in solids can have practical application as a way to obtain structures of given shape and size. In the PbTe/CdTe immiscible material system thermal annealing of two-dimensional PbTe epilayers embedded in a CdTe matrix leads to a creation of PbTe quantum dots \cite{Groiss1,Groiss2,Groiss3}. The size and shape of the quantum dots can be controlled by time and temperature of after-growth annealing processes. The case of morphological transformations in PbTe/CdTe multilayer heterostuctures was studied experimentally in Ref.\onlinecite{Karczewski1}. It was shown that depending on the growth conditions structures of different dimensionality were formed. Only the structures grown at the   lowest temperatures  preserve the intended, two-dimensional, layered  structure.  When the temperature  is increased  the layers transform into zero-dimensional dots that then  merge creating    one-dimensional columns. At sufficiently high growth temperature   PbTe and CdTe deposits separate completely forming thick, quasi bulk objects\cite{Karczewski1}. 
Self-organization  by spinodal decomposition  in solids is  interesting not only from the point of view of the practical applications but also from the point of view of  theoretical study of processes leading to morphological reorganization of deposited structures.  Detailed study of these processes allows for deeper understanding and better controlling of ordering, dimensionality, shape and dimensions of nano-material systems depending on differently prepared initial states and various thermal conditions during the formation process.  

Different aspects  of spinodal decomposition that  happen in semiconducting materials  have been  studied  up to now\cite{dietl,Groiss1,designMC,mouton, rosenthal,kawasaki,sato}. In Ref.\onlinecite{Groiss1} Cahn-Hiliard equation was applied to the description of the experimental results. It reconstructs a pinhole creation evolving  then into a quasi-one dimensional structures elongated in the plane of layer and finally  transforming into a drop pattern.  This  kind of description does not include behavior of multi-layer structures.  Formation of column structure within multi--layered system in most studies appears as a result of surface diffusion.  In Ref.\onlinecite{designMC} it is shown that two-dimensional spinodal decomposition under layer by layer crystal growth condition leads to quasi one-dimensional nanostructures  -- so called  Konbu-phase. Within the same picture bulk diffusion without layer by layer growth leads to the three dimensional  structure  of small crystals.  Rosenthal et. al \cite{rosenthal} studied  self-organization process  via surface diffusion of magnetic clusters in metal- polymer nanocomposites.  Surface-mediated  self-assembly of ErSb/GaSb nanocomposite structures was proposed in Ref.\onlinecite{kawasaki}.   Surface diffusion formation mechanism of  a column-like nanostructures was also shown in  Mouton et. al. study \cite{mouton} of binary alloys.  A hybrid model of  ab-initio and MC numerical calculation  was studied in Ref.\onlinecite{sato} leading again to isotropic, three-dimensional  phase called dairiseki.
All mentioned above studies of self-organizing processes result in structures that build uniformly in all sample at once, or as in the case of layer by layer growth, they begin at the surface and then stay homogeneous top to bottom of the system.
 In this paper we propose the model that accomplishes  bulk and surface diffusion, as well as, it takes into account different mobility of material components forming the superlattices and the existing  temperature gradients. We study spinodal decomposition process in the system  by kinetic Monte Carlo simulations of the model at different temperatures and growth conditions. The characteristic feature of the process we study is that the elongated structures build in time and as an effect in the space, beginning from the first grown, bottom multilayer. As an effect we obtain characteristic structure of gradually changing morphology of the sample from bottom to top.
 The model explains within one picture all phenomena that are observed in the growth of  PbTe/CdTe superlattices.   We discuss the importance of the given mechanism in the observed process of creation of zero--,  one--, two-- and finally three--dimensional structures.

\section{Structure formation}
\begin{figure}
\includegraphics[width=7.5cm]{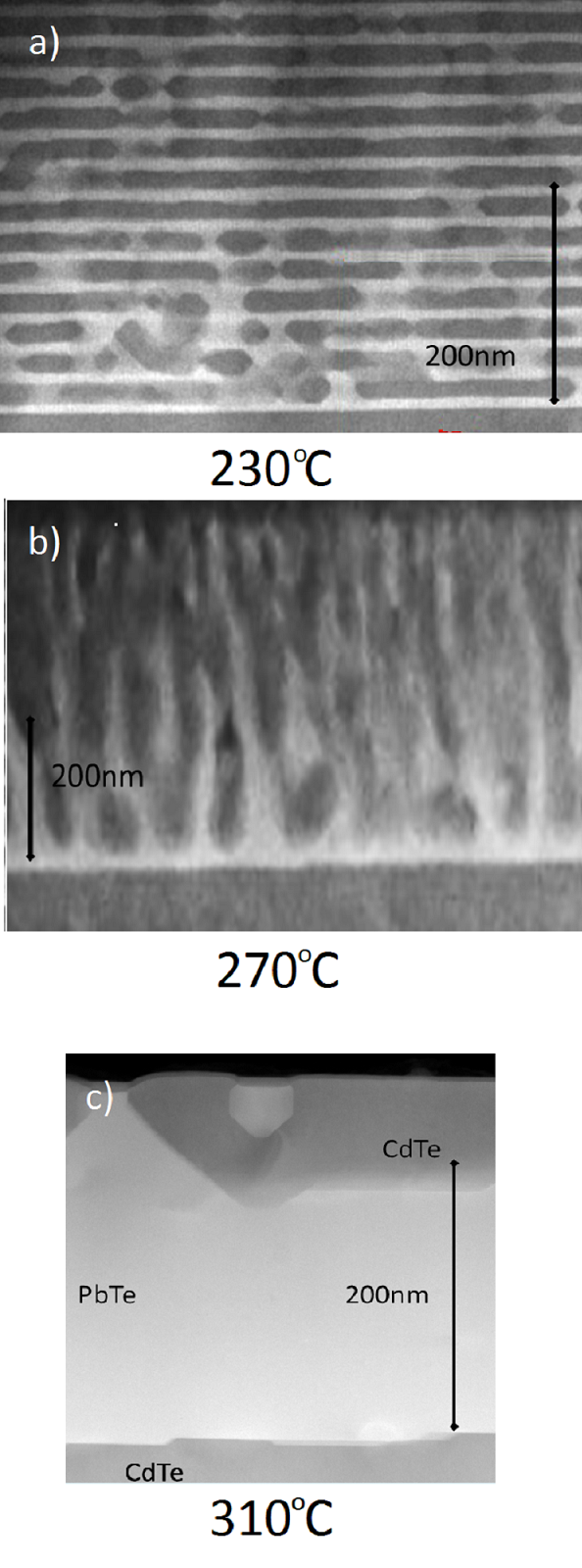}
\caption{Scanning transmission electron microscopy (STEM) image of PbTe/CdTe multilayer structure grown at a) 230$^o$C b) 270$^oC$ and c) 310$^o$C. The growth proceeded from bottom to top.  CdTe structure is plotted in dark gray and PbTe in light gray as noted in c)}
\label{exp}
\end{figure}

The structures described in Ref. \onlinecite{Karczewski1}   were grown using molecular beam epitaxy at three different temperatures: 230\degree C, 270\degree C and 310 \degree C. The growth process consisted of alternate deposition of PbTe and CdTe layers, each for 60s and 180s, respectively. The whole process lasted for 25 such repetitions. In  Fig. \ref{exp} we show scanning transmission electron microscopy (STEM) images of  structures
grown at three different temperatures. Ref. \onlinecite{Karczewski1} contains detailed analysis of experimental results   for these structures.  We can see that 
only the sample that was grown at the lowest temperature (230\degree C) preserved its two-dimensional, layered character. The average thickness of PbTe layers was found to be 8 nm and that of CdTe layers 21.1 nm. However, the structure obtained at the middle temperature (270 \degree C) was completely different morphologically. Instead of horizontally oriented  layers, vertical one-dimensional, column-like objects of PbTe embedded in CdTe were observed. Finally, at the highest temperature (310 \degree C) both constituents were completely separated, with CdTe aggregated at the top of the PbTe deposit. Moreover significant part of the deposited CdTe had desorbed from the surface of the structure .

This experiment had been motivated by an earlier study by Groiss et al \cite{Groiss1}, which described topological transformation of a single 2D PbTe layer stacked between two CdTe layers, at first into a 1D percolation network, which subsequently disintegrates into islands and then 0D quantum dots. According to the Cahn-Hilliard model the characteristic sizes of those structures can be described by a simple relation $z=(3-D)w$, where $D$ is the dimensionality of the structure and $w$ is the initial width of the 2D PbTe layer.

\section{The model}

\begin{figure}
\begin{center}
\includegraphics[width=9cm]{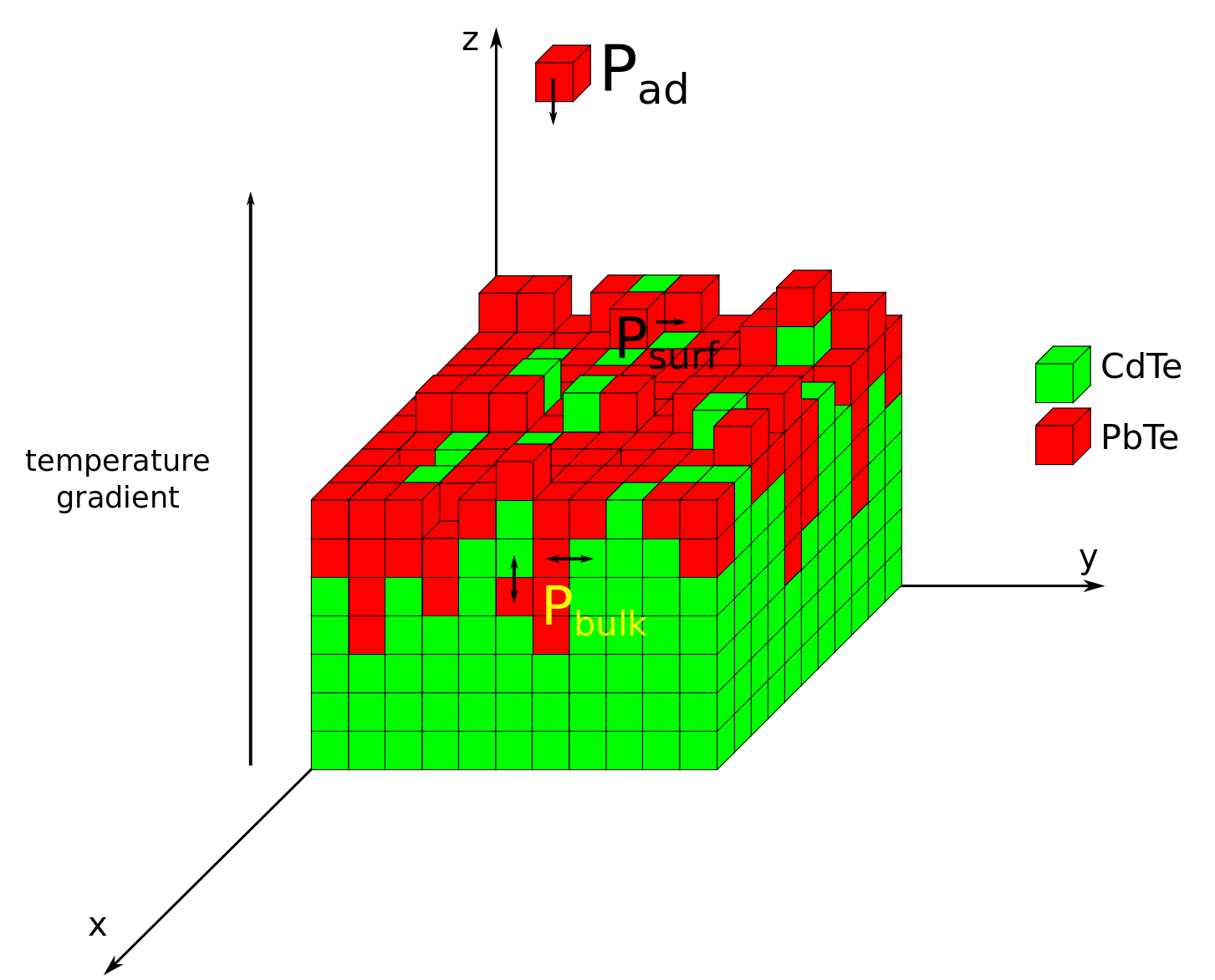}
\end{center}
\caption{Schema of the microscopic model used for kinetic Monte Carlo simulations.}
\label{model}
\end{figure}

Growth of crystal layers is simulated in the kinetic Monte Carlo (kMC) process of a simple 3D lattice  model.  It takes into account three dynamical processes: adsorption of PbTe and CdTe particles at the sample surface, surface diffusion and bulk diffusion of the particles.  The  model is schematically illustrated in Fig. \ref{model}. In each simulation step all  processes are realized  sequentially. At first  new particles are adsorbed at the surface, then they can diffuse on the surface, perpendicularly to the growth direction. At the same kMC step  those particles which are located within the bulk of the crystal may also change their positions. This process, the bulk diffusion,  is isotropic  in the directions  perpendicular to the growth direction. In the direction parallel to the growth axis, however, the bulk diffusion in driven by an additional bias, associated with  the growth course   and  temperature gradient. This sequence of dynamical processes is repeated in every step of the kMC process. We show that  the assumed   set  of events  leads to  the formation  of the structures observed in the experiments. 

Because of the different crystal structure of  PbTe and CdTe (rock salt and zinc blende, respectively) this material system is immiscible, i.e. in its ground state   both components are separated. Such situation can be described as follows: PbTe crystal and CdTe crystals have  higher binding  energy as a bulk than  the interface energy between both materials.  The energy of  the  interface  between PbTe and CdTe slabs 
 have been calculated in  Ref. \onlinecite{Leitsmann1} by the formula
\begin{equation}
E_{int}=\frac{1}{2} \Big{(}E^{tot}[{\rm CdTe/PbTe}]-\frac{1}{2}E^{tot}[{\rm CdTe}]-\frac{1}{2}E^{tot}[{\rm PbTe}] \Big{)}
\label{interface}
\end{equation}
 Where  $E^{tot}[ ~ ]$ means energy calculated for the supercell described within square brackets inside. [CdTe] or [PbTe] refers to  CdTe  or PbTe  supercell  
and [CdTe/PbTe] to  the  supercell with the  interface inside. It was shown in Refs \onlinecite{Leitsmann1,Leitsmann3,bukala} that calculated interface energies are close to  12.5meV/$\AA^2$ and similar   for  all  investigated orientations  (110), (100) and (111). The  formula (\ref{interface}) measures energy cost for creation of the interface between studied nanocrystals  and it can be described as the difference of  the mean energy of bonds between cells  of the same type and  of two different types. These energies decide about the final shapes of the structures that are  built in the  decomposition process of two systems. We use these results in modeling of  the energy  of the arbitrary state of the system.  Let us make coarse grain description of the system. Our lattice is the simplest, cubic one and each site represents one unit of PbTe or CdTe crystal of size $a$.  The energy of the system at given state can be described as
\begin {eqnarray}
H={\rm x}_{\rm PbTe}E^{tot}[{\rm PbTe}]+{\rm x}_{\rm CdTe}E^{tot}[{\rm CdTe}]+\nonumber \\
J_{nn}\sum_{<i;j>}^{nn} n^{i}_{\rm PbTe}n^{j}_{\rm CdTe}  +J_{nnn}\sum_{<i;j>}^{nnn}n^{i}_{\rm PbTe}n^{j}_{\rm CdTe},
\label{en}
\end{eqnarray}
  where  x$_{\rm CdTe}$, x$_{\rm PbTe}$ denote fractions of the given compound  in the whole system, $nn$ and $nnn$ mean the nearest and next nearest neighbors of different types.  The occupation of the $i$-th site by a cell containing a PbTe(CdTe) pair $n^{i}_{PbTe(CdTe)}=0;1$.  The bond  parameters $J_{nn}$ and $J_{nnn}$  attributed to  [CdTe/PbTe] pairs 
in this approach measure  the   energy change in the system due to creation of such pair. With the following values  $J_{nn}=13meV$ and $J_{nnn}=40 meV$ the interface values from Refs \onlinecite{Leitsmann1,Leitsmann3,bukala} are reproduced. All interface orientations (111), (100) and (110) have similar values. As it will be shown later as a result  the shape of the formed dots made of one of the components is similar to that observed in experiments\cite{bukala,Heiss}. Note that in order to calculate the  interface energy or the difference between energies of two system  states the values of $J_{nn}$ and $J_{nnn}$ only are  enough. All other contributions to the energy cancel out.

We start our simulations with  a thick CdTe layer located at the bottom of the structure. The layer represents the 4$\mu$m thick CdTe buffer located between the substrate and the PbTe/CdTe structure in the real experiment. In the experiment the thickness of the buffer layer is much larger than that of the deposited layers, which have the thickness of several nanometers. Because of computational limitations, however, the buffer in our simulations is only 50 atomic layers thick.
We start MC simulation by  adsorption of new particles on the initially prepared buffer.  The particles are  adsorbed at randomly chosen sites of  the substrate or later  on  the top of the already deposited particles.  The probability of this process is given by $P_{ad}$. We set  $P_{ad}=2\cdot 10^{-13}$ ML/(MC step), what assuming appropriate time scale is equivalent to a typical, experimental growth rate of 1.3 $\AA/s$.

Particles on top of the crystal and within the bulk of the crystal  structure that is built up can diffuse at each MC step. Two types of diffusion process are present  - the  bulk  and the surface diffusion.  Both are thermally activated, they depend on the energy difference between the initial and the  final state and  there exists an energetic barrier for a single jump of a particle between the lattice sites. 
For the bulk diffusion we do not assume any specific mechanisms. It can be vacancy assisted or happen via interstitial positions. 
All these processes effectively give  different energy barriers and can be projected into the  switching of  the positions of two  nearest neighbors of different type. We realize this process by  choosing  randomly a particle in the bulk and then we select, also randomly, one of its six nearest neighbors. If the particles are of different type, the exchange is realized with the jump rate
\begin{equation}
P_{bulk}=\nu_0 \exp\left(-\frac{B}{k_{B}T}\right)\times\left\{\begin{array}{ccc}\exp\left(-\frac{\Delta E}{k_{B}T}\right)&\rm{for}&\Delta E \geq0\\
 1&\rm{for}&\Delta E <0\end{array}\right.,
 \label{prob}
\end{equation}
where $\Delta E$ is the energy difference of the system related to the change of the neighbors of the involved particles due to their jump calculated with the help of Eq. (\ref{en}). Attempt frequency $\nu_0$ sets the time scale and is assumed to be equal to $10^{12} s^{-1}$. The general energy barrier for diffusion for each single jump  was estimated by the analysis of  rates of the experimental processes at given temperatures what gave us  the value  $B=760$ meV.

With the energy difference calculated with help of Eq.(\ref{en}) jumps lead to configurations where the number of neighboring particles of the same type is larger. This assures that the components do not mix with each other even at higher temperatures and the  system is immiscible.
The process of  diffusion  defined as above does not differentiate any direction. In the process of the growth, however, the top of the crystal is distinguished as a place where particles attach. Therefore vertical  direction can be different for the particle diffusion. The crystal structure can be slightly different or there can even be more vacancies on going to the top of the multilayer structure. Moreover
the bottom of the crystal is the part where the sample is heated.  If the sample is heated from the bottom,  its temperature should decrease towards the top.  
 We assumed that all these effects  result in faster diffusion along the  direction perpendicular to the substrate. To model this property a small driving force is added to the energy difference $\Delta E_z=E+k(z_1-z_0)/a$ where $z_0, z_1$ denote the initial and the final positions of  one  of  the particles  exchanging positions and   $k$ is  a parameter, $a$ - the distance between successive layers. In such a way   exchange of compounds along $z$ axis up or down becomes easier.  We will show that this bias in the bulk diffusion is  responsible for emergence of structures that are prolonged in the z-direction. The biased diffusion is known as  a reason for a formation of  stripped structures \cite{leung,stripped,evans}.

We assume that some vertical temperature gradient is present. This  gradient is assumed to be not large enough to change noticeably diffusion rate from one position to the other, so we leave the diffusion rate across the layers  unchanged. 
However  both material components are not identical. Not only their lattice structures are different but also  the thermal conductivity of PbTe is much lower than that of CdTe (1.9 and 6.2 W/mK, respectively) and their bulk  energies should differ. When the  income from the  temperature gradient  is calculated   as the first correction to the jump barrier in the exponent of rate (\ref{prob}) it becomes proportional to the bulk energy difference.
 PbTe should be then more strongly influenced by the temperature gradient because it is expected to be larger than in CdTe. The difference between the values of the bias should reflect the asymmetry between the two components that was seen in the experiment at high temperatures when the CdTe deposits gather on the top of quasi bulk PbTe. To model this we assume  parameter $k$   different for PbTe and for CdTe. Thus we  set a small difference between values of $k$,  choosing  the bias values for PbTe  moving up $k_{PbTe}=26$ meV and for CdTe moving up $k_{CdTe}=29$ meV. The bulk diffusion is realized in each  MC step together with the surface diffusion at the top crystal layer.

 The surface diffusion is performed by choosing randomly a particle on the surface and making it move to an empty neighboring site at the same level, i.e., with the same z-coordinates or jump down to an empty place to the lower layer.  
Jumps up are disallowed what additionally  promotes the surface smoothening. The probability of a single surface diffusion jump is
\begin{equation}
P_{surface}=\nu_0 \exp\left(-\frac{B}{k_{B}T}\right)\times\left\{\begin{array}{ccc}\exp\left(-\frac{\Delta E}{k_{B}T}\right)&\rm{for}&\Delta E\geq0\\
 1&\rm{for}&\Delta E<0\end{array}\right..
\end{equation}
The values of $\nu_0$ and $B$ are kept the same as in the process of bulk diffusion, however because particles are less connected  to the surface and they can jump to the empty site the surface processes are faster than bulk ones.
For the studied systems   the role of the surface diffusion  is reduced mainly  to smoothening of  the surface after new particles are adsorbed on it. We will show however that in different conditions  the  process of surface diffusion in our model can lead to the ordered structure of one dimensional stripes as it was discussed before \cite{surf1,surf2,surf3}.

In our kMC simulations subsequent layers are  grown in the similar way as it was done in the experiment. The type of the currently adsorbed particles changes periodically, starting from PbTe. For the fixed value of the adsorption rate  deposition duration  of each of the components can be chosen in such a way that in each repetition PbTe and CdTe consist of the desired number of atomic layers. Assuming the lattice constants $a$ of 6.46$\AA$ for PbTe and 6.48$\AA$ for CdTe, we need approximately 12 atomic layers of PbTe and 32 atomic layers of CdTe to simulate the real growth conditions. It corresponds to the thicknesses of the layers used in  the experiment \cite{Karczewski1}. We have performed Monte Carlo simulations based on the microscopic model described in this section. We set the size of the crystal $N_x=N_y=$80 particles both in the x and y direction. We assume the periodic boundary conditions in both those directions. The maximal size in the z direction (the direction of the crystal growth) is $N_z=300$ particles.
The kinetics of the model is dependent on six material   parameters: $P_{ads}$, $J_{nn}$, $J_{nnn}$, $k_{PbTe}$, $k_{CdTe}$, $B$. The temperature  $T$ and the pattern of particle deposition are external parameters. We  reproduce the different experimental results by changing only the external parameters while keeping all the remaining parameters constant.
\begin{figure}
\includegraphics[width=7cm]{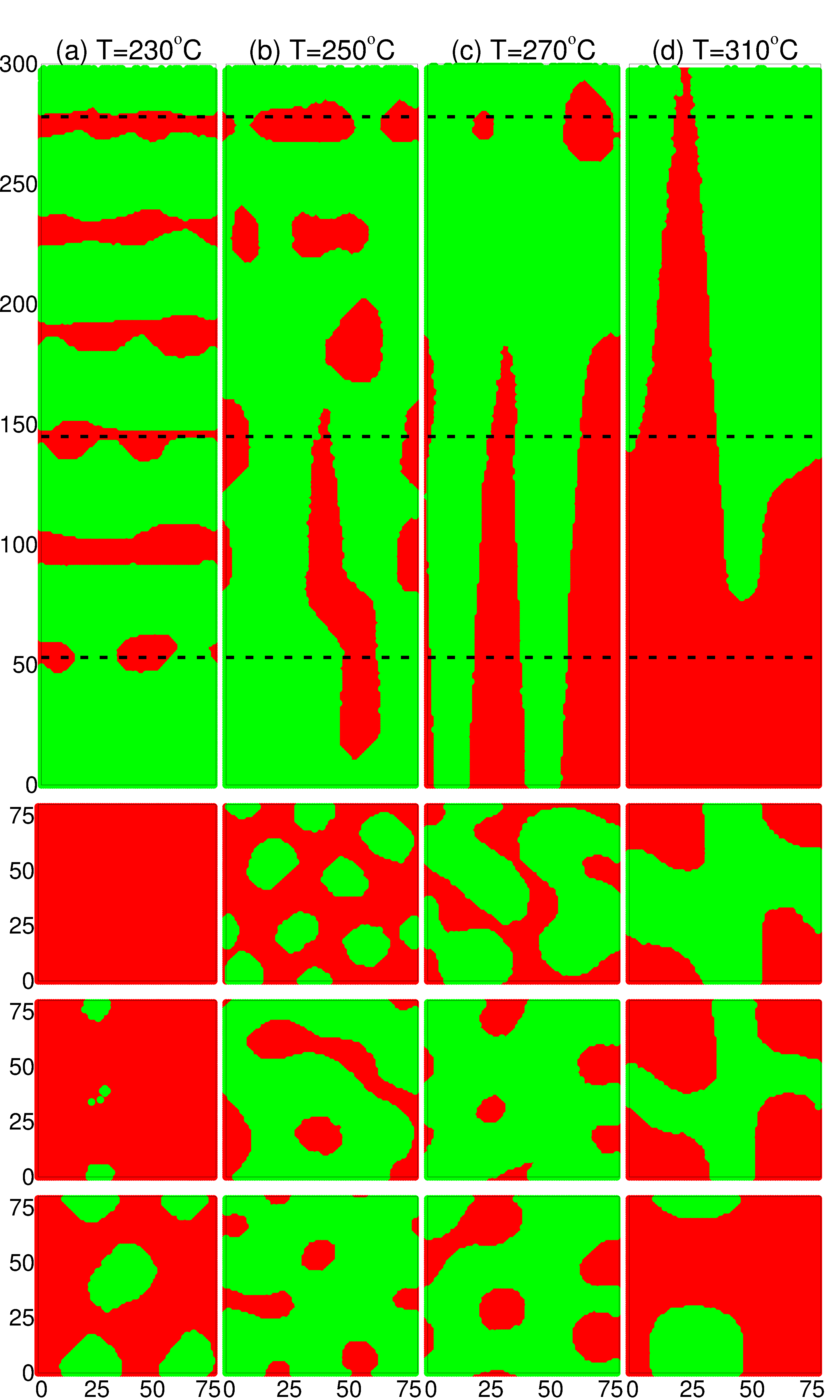}
\caption{ Vertical (top) cross-sections and horizontal (three bottom rows) of different structures that emerge during PbTe/CdTe growth.
 Temperature increases from left to right T=230$\degree$C (a); 250$\degree$C (b); 270$\degree$C (c); 310 $\degree$C (d). Bottom panels represent successive horizontal cross-sections at successive z values  marked by dashed lines in top panel.  }
\label{cross_sections1}
\end{figure}

\section{Results}
\subsection{Simulations with vertically biased bulk diffusion}

By using the MC model described in the previous section we simulated the morphological transformations observed experimentally in PbTe/CdTe material system \cite{Karczewski1}. Below, we study specific situation  of the reorganization process during growth leading to the appearance of zero--,  one--,  two-- or three-- dimensional patterns. With the assumed  values of the material parameters we perform simulations for a few various temperatures $T$. We run it for several repetitions of the deposited layers. The vertical cross-sections, i.e., along the growth direction, obtained as a result of such numerical procedure are shown in upper panel of Fig. \ref{cross_sections1}. In order to see the changes in the simulated structures more clearly the lower panel of  Fig. \ref{cross_sections1} shows horizontal cross-sections of the structures at the levels marked in the upper panel by dashed lines. The growth of the component layers was simulated to the  thicknesses corresponding with the experimental ones. 

At the lowest temperature of  $T=230 \degree $C presented in  Fig \ref{cross_sections1}a it is clearly seen that  six PbTe/CdTe layer  repetitions have been deposited. Above CdTe substrate we can see  full six PbTe layers, full five CdTe layers and a thinner CdTe  cap layer on the top of the structure. As in the experiment, the structure simulated for $T=230 \degree $C exhibits 2D, layered character. However, one can notice that  the PbTe layers deposited earlier, i.e., located closer to the substrate differ from those deposited later, i.e., located closer to the structure surface. For instance, the highest PbTe layer, plotted as  the  first from the top, has no CdTe holes inside. The holes appear, however, in the lower layers and in the lowest one a few CdTe holes of larger size are clearly visible. The observation indicates an onset of the reorganization of horizontal 2D layers into  1D percolation networks, which later would disintegrate into PbTe quantum dots embedded in CdTe matrix. The process much more affects the layers which were deposited first because they stayed at the growth temperature for longer time than those deposited at the end of the growth. This indicates that the bulk diffusion plays the role in the reorganization process. In our simulations the material located deeper in the structure underwent much more diffusion processes. It is interesting to note that our simulations predict that even at such low temperatures as  $T=230 \degree $C the layers undergo morphological transitions. These observations are in agreement with the results of the experiment on the multilayer PbTe/CdTe  structures \cite{Karczewski1}.

The same amount of  PbTe and CdTe materials deposited at  the other temperatures results in structures with quite different morphology as shown in Fig.\ref{cross_sections1}b-d. The layered character is no more  evident or does not exist at all. In Fig. \ref{cross_sections1}c we can see the result of the simulation at the next experimental temperature of $T=270 \degree$C. At the bottom of the sample vertical PbTe columns  have emerged. Above the PbTe columns separated PbTe dots can be seen. Such system ordering agrees perfectly with the experimental results, where cross-sections of the system at this temperature looks very similar. The vertical cross-section of the developed  columns can be seen  in the lower panel of Fig. \ref{cross_sections1}c. The cylindrical  shape is clearly seen and it can be observed that its diameter  is larger than the initial thickness width of the PbTe layers. The diameter increases towards the bottom of the sample. It is an effect of the stronger bias of the vertical bulk difussion for CdTe than for PbTe. In the experiment the PbTe columns also had larger diameter at  the bottom and near the substrate they are  so wide that they begin  to merge. 

Fig. \ref{cross_sections1}b shows the results of  growth  at temperature of $T=250 \degree$C, which is between two temperatures discussed above. It can be seen that the system behaves exactly as it is supposed to. The elongated structures that build at the bottom are shorter than the ones for the higher temperature  and reach only the second and third  layer.  Comparing the cross-sections in Fig. \ref{cross_sections1}b and \ref{cross_sections1}c we can see that the column diameter increases with the increasing temperature. The higher temperature accelerates bulk diffusion what leads to the formation of larger structures.  The layer above consists of droplets and top PbTe multilayers are holed  resembling bottom multilayer in the system at the  lower temperature. 

  Fig. \ref{cross_sections1}d shows the highest experimental temperature $T=310 \degree$C in which we see that PbTe compound is moved down whereas CdTe structure appears above it. Such tendency is also seen in the experiment \cite{Karczewski1} where the same alternate deposition of PbTe and CdTe  layers was carried out at $T=310\degree$.  Eventually  two constituents migrated forming large  PbTe layer between the base CdTe   below   and thin  CdTe layer on top. In the simulated system initial  width of the  CdTe base was too small  and as  an effect
all of it  went on top of the system. Nevertheless the tendency of material separation is clearly visible at this temperature.  Division of two compounds into bottom and top layers in our model is caused by the small difference in the bias for particle interchanges.

\begin{figure}
\begin{center}
\includegraphics[width=8.5cm,angle=0]{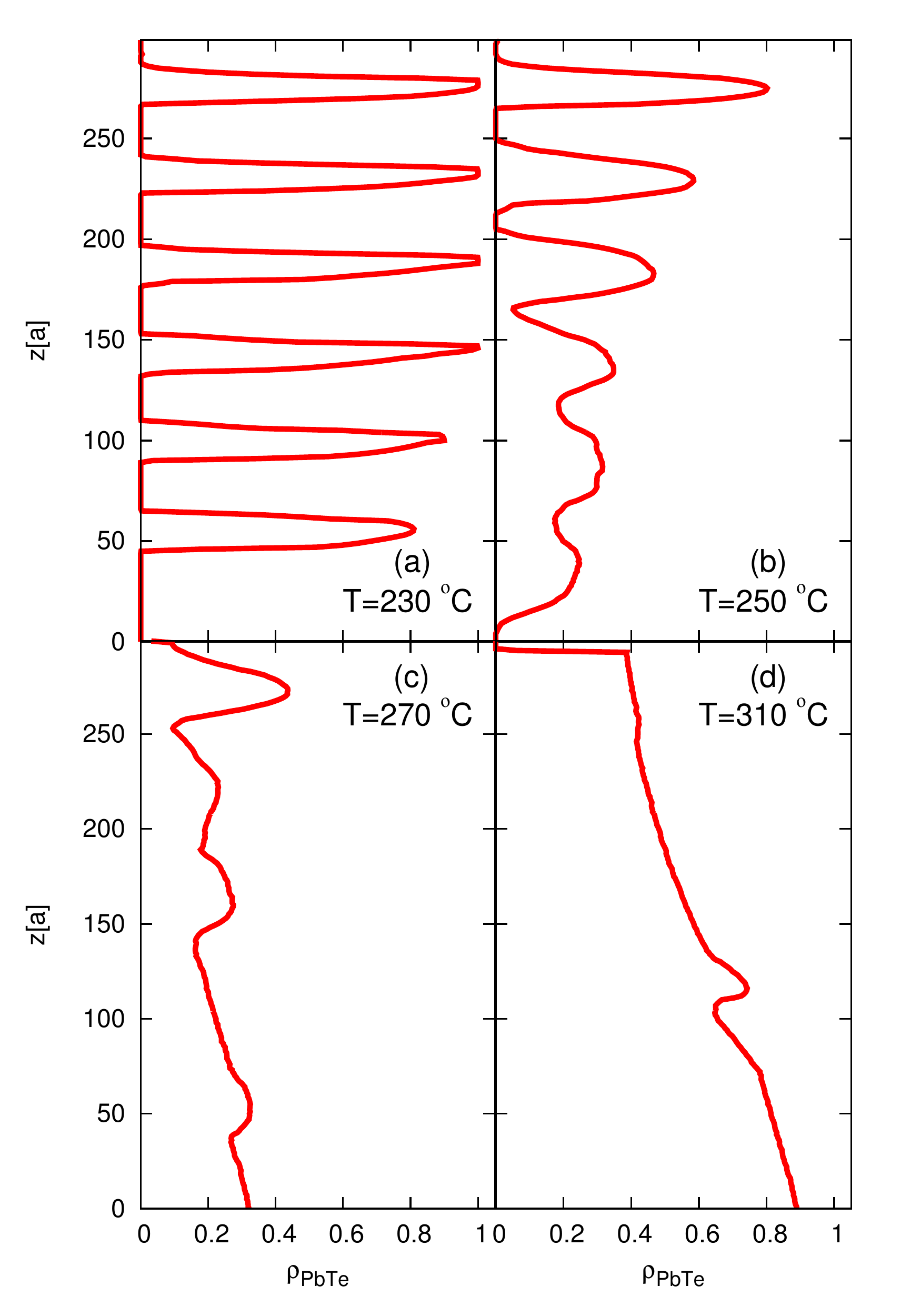}
\end{center}
\caption{PbTe density along z axis at temperatures 230$\degree$C (a); 250$\degree$C (b); 270$\degree$C (c); 310$\degree$C (d). }
\label{cor}
\end{figure}

\begin{figure}
\begin{center}
\includegraphics[width=6cm,angle=-90]{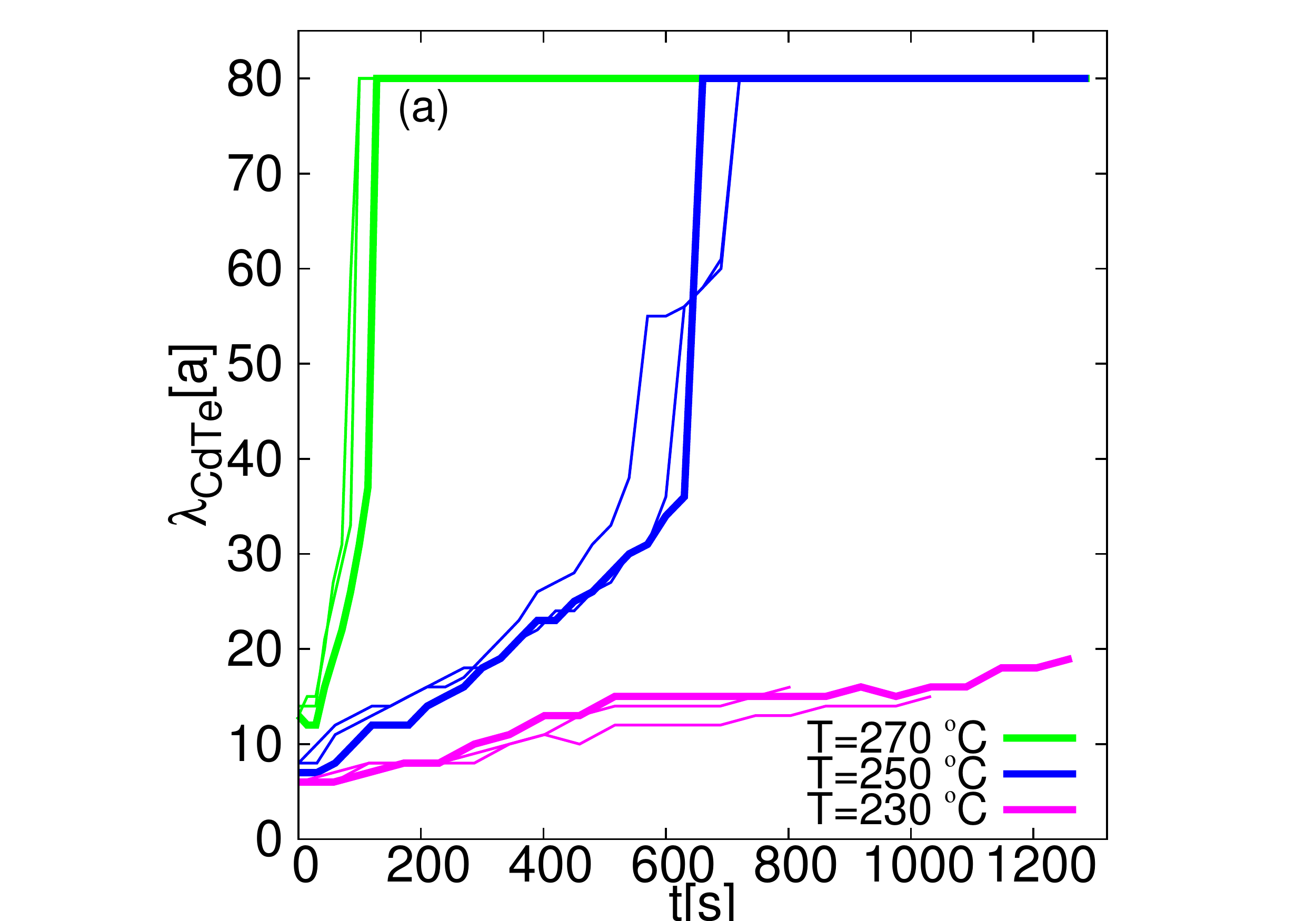}
\includegraphics[width=6cm,angle=-90]{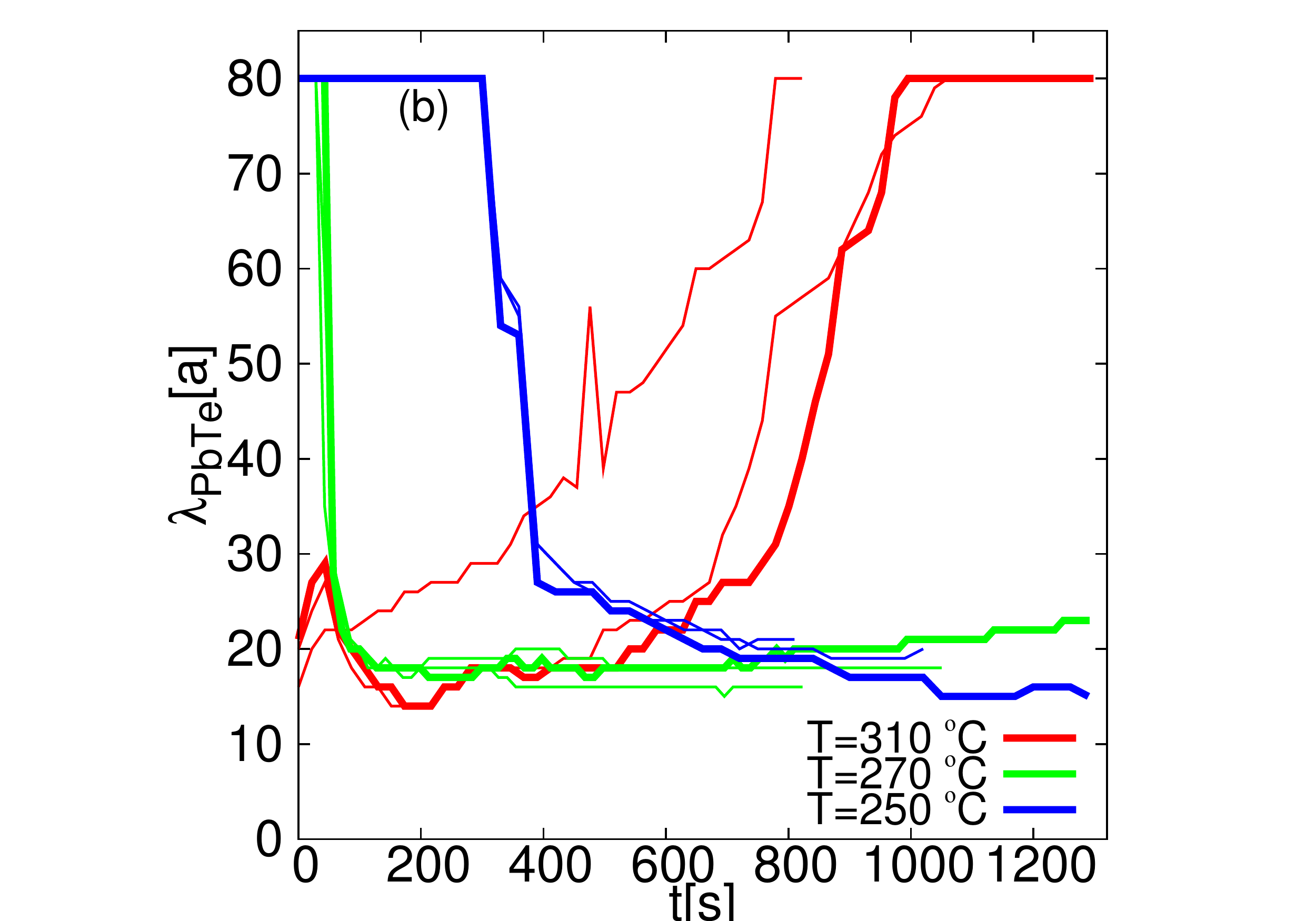}
\end{center}
\caption{(a) Correlation length  $\lambda_{CdTe}$ in vertical direction for systems grown in  T=230$\degree$C; 250$\degree$C; 270$\degree$C  and  (b) correlation length  $\lambda_{PbTe}$ in vertical direction for systems grown in  T=250$\degree$C; 270$\degree$C; 310$\degree$C as a function of time $t-t_0$. Three lines represent three consecutive layers from bottom to top. 
For each layer time $t_0$ was  set as a  time when deposition of given layer started. The most thick line means the first, bottom layer in each case. }
\label{corr2}
\end{figure}

The configuration of the grown system as a function of the height can be seen in Fig. \ref{cor} where the particle density at given height   $z$ 
\begin{equation}
\rho^{z}_{PbTe}(r,z_{0},t)=\frac{1}{N_x N_y } \sum_{x,y} \phi_{PbTe}(x,y,z)
\end{equation}
 is plotted, where  $\phi_{PbTe}(x,y,z,t)$  is  1 for PbTe particles and 0 otherwise. We sum over all  $x$ and $y$ positions, whereas $z$ is fixed.  We can see in Fig. \ref{cor} that at the lowest temperature of $T=230^oC$ all layers are well separated. Moreover it can be noticed that  only the first and the  second layer at the bottom have holes, whereas higher layers are slightly deformed but not perforated because density reaches value 1 for all of them. For higher,  $T=250 \degree$C we  see three separated layers at top, and the structures that extend to three successive bottom  layers. At the higher temperature $T=270 \degree$C only the top layer preserves but its height is reduced, what means that it has  mixed   structure.   And finally we have the highest  temperature studied $T=310 \degree$ C, where neither in the density plot in Fig \ref{cor}, nor in the cross-sections  striped or  dotted pattern can be seen. Instead we have perfect separation of the system into two phases - one PbTe at the bottom and the second CdTe at top. 

The structures shown  in Fig. \ref{cross_sections1} and \ref{cor} are created in the process of bulk diffusion. Note, that here, in immiscible compounds bulk diffusion leads rather to separation than  particle mixing. Because  this process happens  all the time during layer deposition, thus  structures seen  top to bottom of grown multilayer structure  represent  consecutive moments of time evolution. Time behavior of the formed structures  can be analyzed more precisely by the 
following   correlation function  
\begin{align}
C_{\alpha}(\vec{r},z,t)&=<\phi_{\alpha}(x,y,z,t)\phi_\alpha(x+r_x,y+r_y,z,t)>
\label{corr}
\end{align}
where $\alpha$ can be PbTe or CdTe. The correlation defined  above is averaged (what is noted by symbol  $<>$)  over all  positions $x,y$ within each horizontal  layer. It describes occupation  correlations  at a  distance $\vec r$  at given height $z$ and time $t$.  The model is  isotropic in the $x y$  directions hence we will describe all relations further as a function of length $r=|\vec r|$.
 Let us take  $C_{\rm PbTe}$. For $r=0$ it is equal to the number of PbTe particles in the chosen layer at time $t$. For increasing   $ r$  $C_{\rm PbTe}$  stays at the same level as long as each neighbor is PbTe and it decays  if a CdTe particle   is present at a given distance. After calculating the  average (\ref{corr})  the obtained correlation function decreases as a function of distance as  $C_{\rm PbTe}=C_0 \exp(- r/\lambda_{\rm PbTe})$. The correlation length  $\lambda_{\rm PbTe}$  is  function of  time $t$ and  height $z$. As an effect  $\lambda_{\rm PbTe}$ measures  time and position  changes  of  the  mean length  of  PbTe structures inside the CdTe. The same we can say  for  CdTe structures by calculating $\lambda_{\rm CdTe}$. 

 In Fig. \ref{corr2} we have shown  $\lambda_{\rm PbTe}$ and $\lambda_{\rm CdTe}$ as a function of time for three consecutive PbTe layers from bottom to top. In any  case time was shifted 
so $t=0$ means  the moment when given layer was completed. 
In the left panel of Fig. \ref{corr2} $\lambda_{\rm CdTe}$ is plotted at three different temperatures. The lowest three lines represent evolution of  layers during  multilayer deposition process at $T=230 \degree$C. The line for the first, bottom PbTe layer is plotted in thicker line. It can be seen that all lines follow the same path. At the lowest temperature they start from the low value, then CdTe holes inside PbTe layer  grow up to the mean size of  20$a$. In Fig.\ref{cross_sections1}a we can see that holes of bottom layer  really have diameters of such size. At the higher temperatures $T=250 \degree$C  and $T=270 \degree$C CdTe structures grow quickly and eventually reach size of the system of 80 lattice sites.  At the same temperatures  PbTe structures decay coming to value 15 lattice sites at $T=250 \degree$C
and 22 lattice sites at $T=270 \degree$C. These numbers describe mean diameter of the final PbTe structures. Note that all lines follow the same shape. Such property confirms hypothesis that structures inside system are build  via bulk diffusion what happens all  time during deposition process. Moreover higher temperature leads to slightly larger structures.
Evolution of the system at the  highest T=310$\degree$C is illustrated in Fig \ref{corr2}b, where correlation function for PbTe crystalic structure is shown. At the topmost temperature PbTe structure starts from the width of 20 lattice sites  and then grows, eventually reaching the size of the system. Differently than all other curves at this temperature all three lines do not follow each other. It can be seen
that the system reorganizes as a whole forgetting about layers just from the beginning.

\begin{figure}
\begin{center}
\includegraphics[width=8cm, angle=-90]{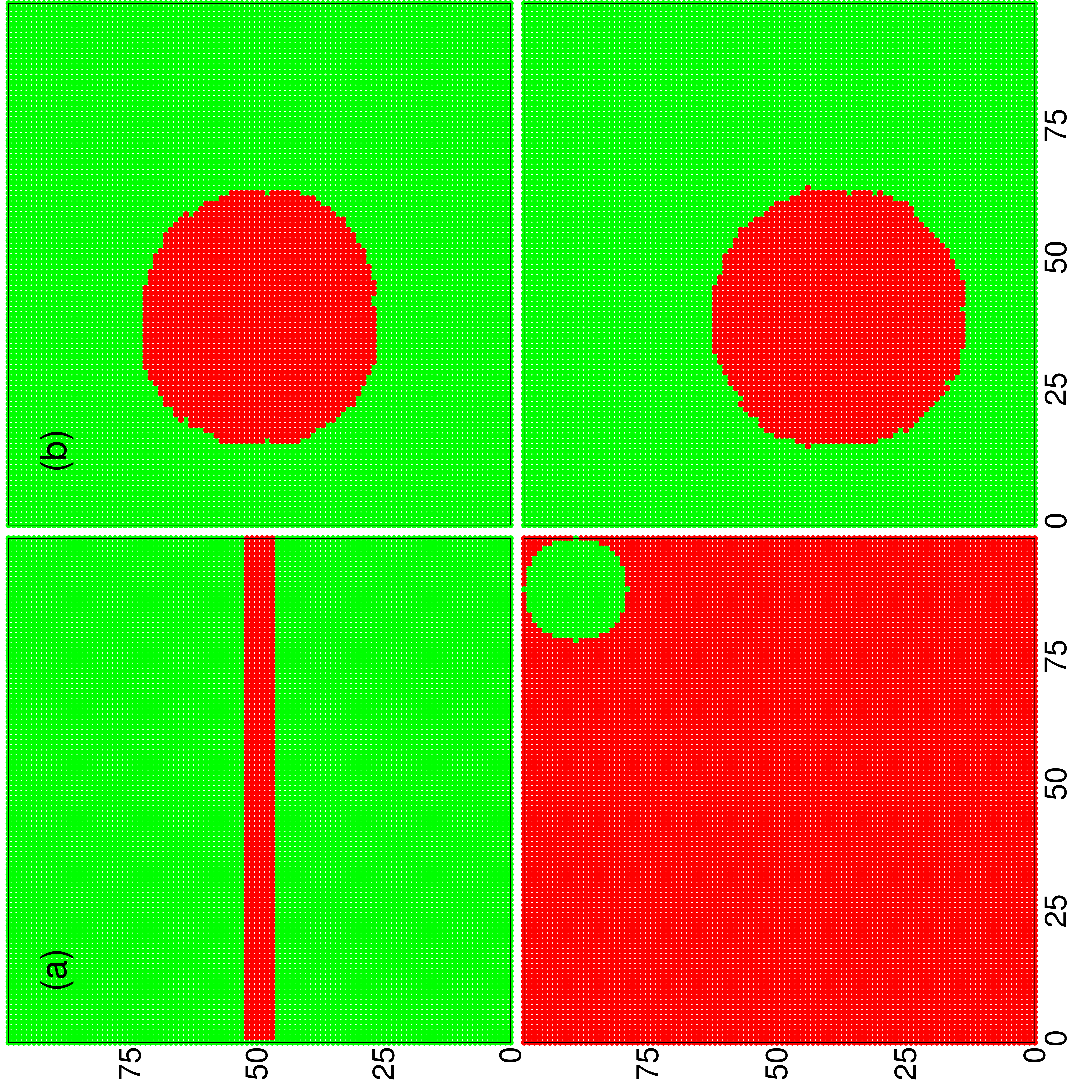}
\end{center}
\caption{Transformation in the one-layer annealed system at $T=350 \degree$C. Left side (a) present initial situation and right side (b) the annealed system.  Top panels show  vertical cut  and  bottom panels show horizontal  cut. }
\label{Groiss1}
\end{figure}
\subsection{The role of  bulk diffusion}
We have shown that biased model reproduces all experimentally observed types of system evolution at different temperatures. The structure with holes inside layer transforms  into the percolating structure  and  finally to the spherical droplets. Above mentioned structures are   formed due to the presence of  isotropic bulk diffusion, whereas columns build up  because of   bias. The bias  is related to  the anisotropies present within growing multilayer structure together with the temperature gradient which additionally responses  for the up down anisotropy of PbTe/CdTe interchange in vertical direction.   One can expect, thus, that in a system evolving without any anisotropies, for instance,  a single PbTe layer embedded in thick CdTe bulk crystal annealed by isotropic heating final structures of  spherical shapes should be found. 
\begin{figure}
\begin{center}
\includegraphics[width=8cm, angle=-90]{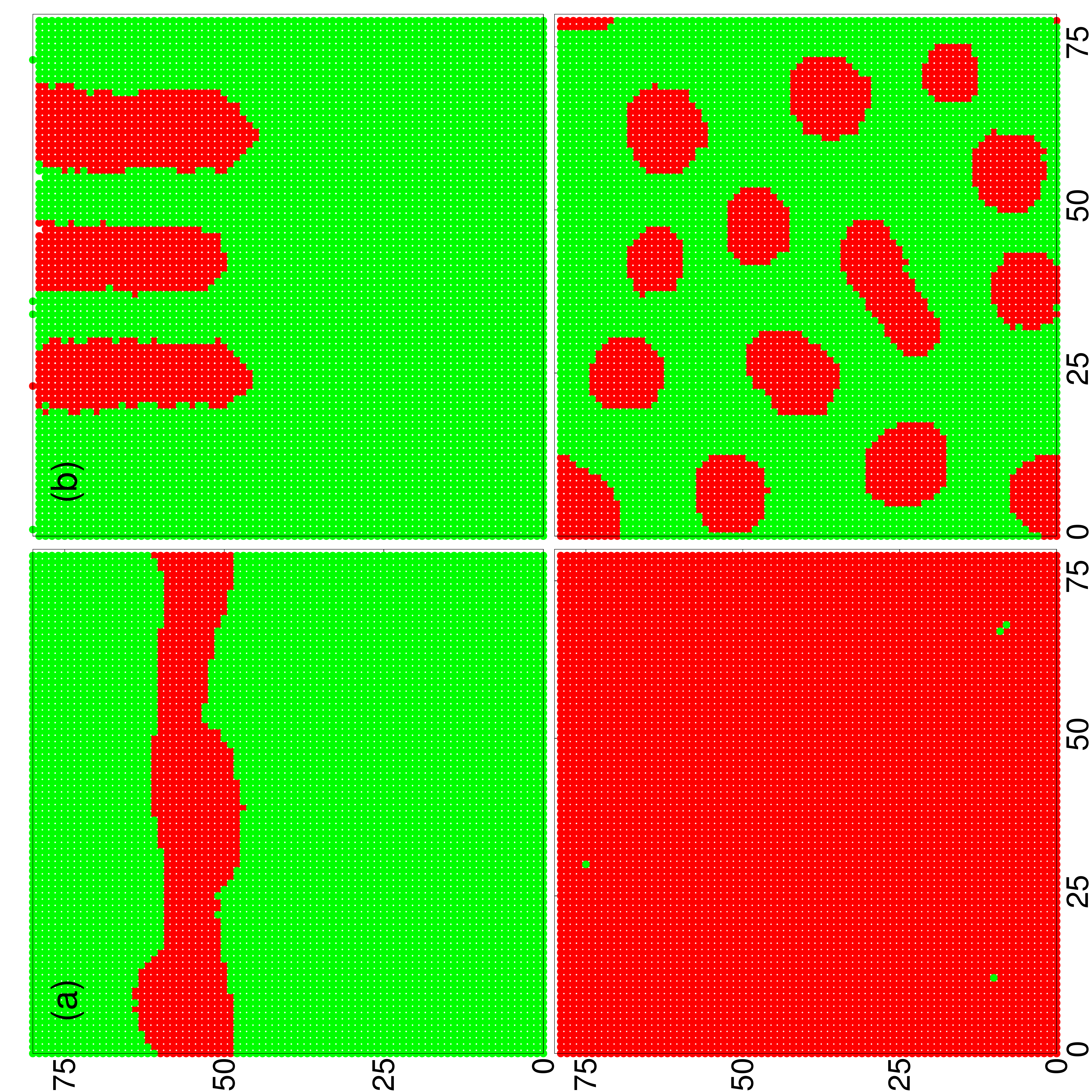}
\end{center}
\caption{Comparison of two different layering  patterns at low, T= 230 $\degree$C. Left panel (a) shows result of growing layers in order 12PbTe/36CdTe  and  right panel (b) shows the result of growing 1PbTe/3CdTe layers. Top panels are plotted for vertical an bottom panels for vertical cross-sections. All other parameters, temperature, mean  growth rate and the model constants are the same. }
\label{layers}
\end{figure}

Such behavior was indeed observed in the experimental systems \cite{Groiss1,Groiss2,Groiss3} and to compare results we run our simulations at the experimental temperature of $T=350\degree C$  without any bias. In the small system we study without any bias the  surface tension  appeared to be  strong, so to initiate the process we had to cut a single hole within initially prepared  layer.   Experimentally such holes were observed at the beginning of the annealing process \cite{Groiss3}. Initial ordering of the system is shown in the left panel of Fig \ref{Groiss1} where six parallel layers of PbTe are embedded in the CdTe  surrounding. A single hole  of a diameter of 20 lattice units filled out by CdTe was present from the beginning. This choice reflects the size and density of experimentally measured structures.
 Result after 13 minutes of annealing  at $T=350 \degree$C can be seen in Fig. \ref{Groiss1}. Because of the character of this process only bulk diffusion participated in the transformation. There was no open surface of PbTe in the annealed system and  as an effect the surface diffusion does not occur at all.  The initial structure was 2D with a small cylindric hole, then we observed gradual growth of this hole and change of the layer. Finally perfectly spherical shape of resulting PbTe structure should be noted. We  have not seen elongated 1D structures. They are expected in the case when changes are initiated by several holes and elongated structures build between them.
\begin{figure}
\begin{center}
\includegraphics[width=6cm,angle=0]{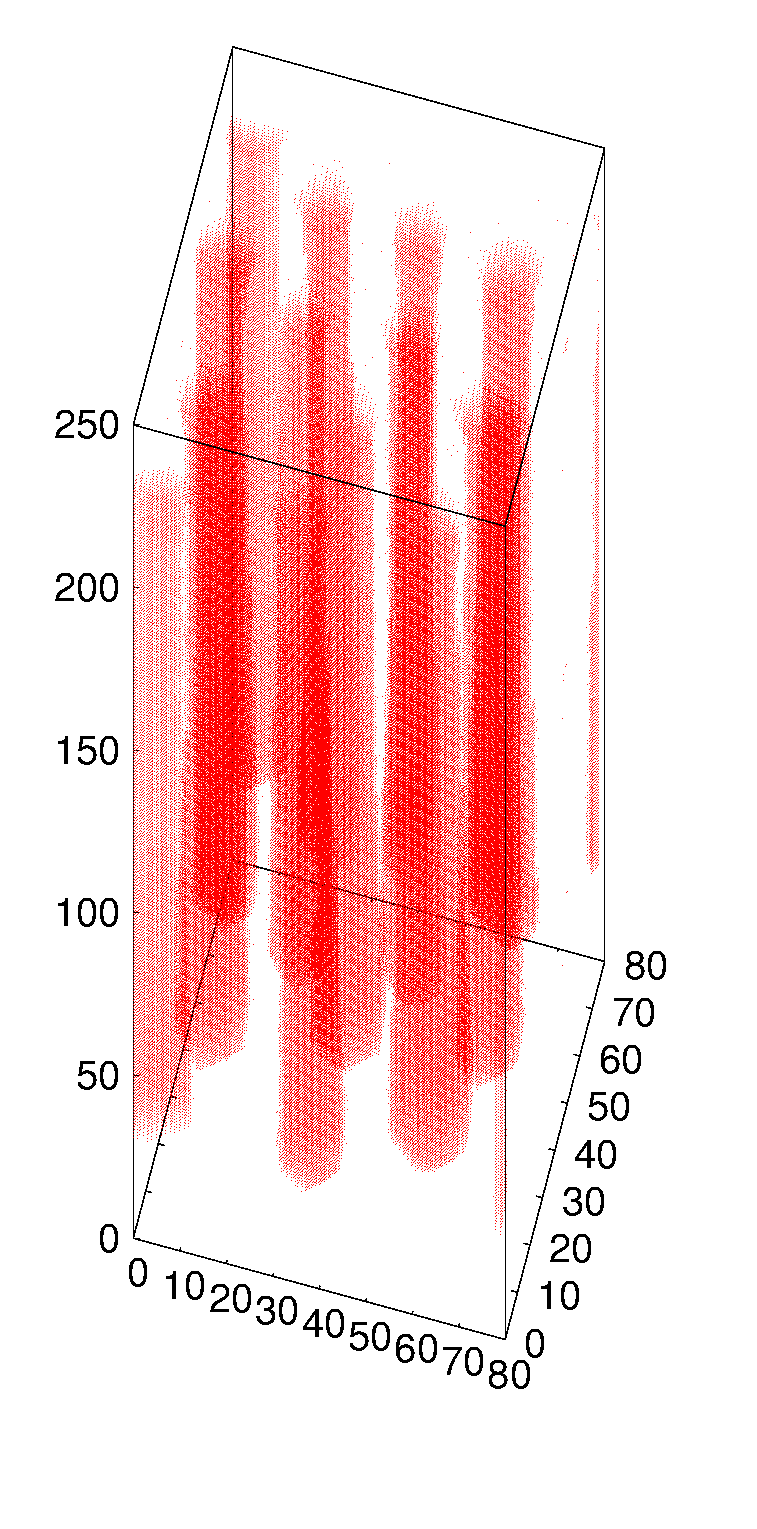}
\end{center}
\caption{Results of surface diffusion obtained at 230$\degree$C by growing 1PbTe/3cdTe layers.}
\label{3D}
\end{figure}
\subsection{The role of the surface diffusion}
The vertical cross-sections in Fig. \ref{cross_sections1} show that the layers deposited first, i.e., located deeper in the structure, differ from those deposited later. The same observation is confirmed by  data shown in Figs. \ref{cor} and \ref{corr2}. The morphology of the structures changes in the vertical direction. This effect indicates that  bulk diffusion is the most important factor which determines the final configuration of the structure. Yet the surface diffusion takes also place in the self-organizing  process and its mean rate is of the same order as this of the bulk one. The main role of the surface diffusion in our system is the surface alignment.
Yet, many works  show how the  stripe ordering results as an effect of  the surface diffusion\cite{dietl,designMC, mouton, rosenthal,kawasaki}. Such process indeed is possible in our system also  but not at the experimental conditions that are studied here. To illustrate this we  show that  the deposition of thinner multilayers eventually  leads to the stripe formation starting at the surface of grown crystal. We studied the process of crystal grow at the lowest temperature of T=$230 \degree$C  and with the same growth rate as in Fig. \ref{cross_sections1} and the ratio of delivered components is the same. What we have changed as compared to the first situation is the  width of the deposited multilayers. We have grown the crystal by setting one PbTe layer and three CdTe layers. In Fig~\ref{layers} we can see that the geometries of the grown structures are completely different in left panels - for thicker layers and in right panels - for thinner layers. Thicker multilayers stay more or less two-dimensional, they eventually  perforate during further growth process. In the second case, when multilayers are very thin striped structure builds just from the beginning. The main mechanism of this process is surface diffusion, that has a possibility of ordering the material during growth process. The final striped structure can be seen in Fig~\ref{3D}. Resulting  columns  are straight, spread from bottom to top, but they have some additional structure in the vertical direction. They are wider at the bottom than at the top. This last feature evidently is due to still active bulk diffusion and the existing anisotropy in the component diffusion. When surface diffusion is left as the only dynamical process in the system above it results in creating columns very similar to this seen in Fig.~\ref{layers}, but not so regular,   with not so smooth surfaces and of the same width from bottom to top.

\section{Conclusions}
We  proposed model that  correctly reproduces all types of experimentally observed morphological evolutions of the PbTe/CdTe structures at different temperatures\cite{Karczewski1}. Depending on the growth conditions  combination of   bulk and surface diffusion cooperate in the process of spinodal decomposition of immiscible materials and lead to the formation of 1D elongated structures or 0D spherical droplets. Additionally,  present in the process of crystal  growth  vertical bias orders created structures in the shape of  vertical columns  collecting material from several deposited layers. Formation of such columns starts from the bottom of multilayer structure and lasts all time until structure is grown. This is exactly  the case of experimental systems \cite{Karczewski1}. The small difference in bias is main ground for  the phase separation locating  PbTe phase at the bottom and CdTe at top at temperatures as high as $310 \degree$C.
We have  shown  that  the same model leads to other patterns like long columns spreading from bottom to to top when different deposition schema is applied.  It can be seen that the structure formation in the immiscible compounds can be controlled both by temperature change and by the deposition method.

\begin{acknowledgments}
The research  was partially supported by   the National Science Centre (NCN) of Poland (Grant NCN No.2015/17/N/ST3/02310) (MM), Grant No. DEC-2014/14/M/ST3/00484 and by 
 the Foundation for Polish Science through the Master program (G.K.).  
\end{acknowledgments}

\end{document}